\documentclass[a4paper,11pt]{article}
\pdfoutput=1
\usepackage[width=150mm,top=25mm,bottom=25mm]{geometry}

\usepackage{amsmath}
\usepackage{mathtools}
\usepackage{amsthm}
\usepackage{thmtools, thm-restate}
\usepackage{physics}    %
\usepackage{amssymb}
\usepackage{amsfonts}
\usepackage{mathdots}   %
\usepackage{mathrsfs}   %
\usepackage{nicefrac}   %
\usepackage{stmaryrd}
\usepackage{tikz-cd}     %

\usepackage{authblk}

\usepackage{xcolor}

\usepackage{xspace}

\usepackage{enumitem}
\setlist{noitemsep}
\usepackage{tikz}
\usetikzlibrary{decorations.markings}
\usepackage{tikzit}
\usepackage{float}

\newcommand{\arrowangle}{142}
\newcommand{\arrowscaling}{1}

\definecolor{dark_grey}{RGB}{70,70,70}
\definecolor{zx_grey}{RGB}{211,211,211}
\definecolor{zx_grey_thick}{RGB}{180,180,180}
\definecolor{zx_red}{RGB}{232,165,165}
\definecolor{zx_green}{RGB}{216,248,216}

\pgfdeclarelayer{backlayer} 
\pgfdeclarelayer{labeltextlayer} 
\pgfdeclarelayer{frontlayer} 
\pgfsetlayers{backlayer,main,edgelayer,nodelayer,labeltextlayer,frontlayer}  %

\makeatletter
\pgfkeys{%
  /tikz/on layer/.code={
    \pgfonlayer{#1}\begingroup
    \aftergroup\endpgfonlayer
    \aftergroup\endgroup
  },
  /tikz/node on layer/.code={
    \gdef\node@@on@layer{%
      \setbox\tikz@tempbox=\hbox\bgroup\pgfonlayer{#1}\unhbox\tikz@tempbox\endpgfonlayer\egroup}
    \aftergroup\node@on@layer
  },
  /tikz/end node on layer/.code={
    \endpgfonlayer\endgroup\endgroup
  }
}
\def\node@on@layer{\aftergroup\node@@on@layer}
\makeatother

\usepackage{pgfplots}
\pgfplotsset{compat=newest, compat/show suggested version=false}

\newcommand{\tikzrefsize}[1]{\ifmmode{\scriptscriptstyle{#1}}\else{\tiny{#1}}\fi}

\tikzstyle{wn}=[font={\scriptsize\boldmath}, inner sep=1mm, outer sep=-1.8mm, scale=0.8, tikzit shape=circle, draw=black, fill=black!01, tikzit fill=white, tikzit draw=black, shape=circle, tikzit category=GLA]
\tikzstyle{bn}=[font={\scriptsize\boldmath}, inner sep=1mm, outer sep=-1.8mm, scale=0.8, tikzit shape=circle, draw=black, fill={rgb,255: red,100; green,100; blue,100}, tikzit draw=black, shape=circle, tikzit category=GLA]
\tikzstyle{whad}=[fill={zx_grey}, draw=black, shape=rectangle, tikzit category=ZX, tikzit draw=black, minimum size=5pt, inner sep=1.5pt, scale=1, font={\scriptsize\boldmath}]
\tikzstyle{wphase}=[
  node on layer= frontlayer,%
  drop shadow={blend mode=color burn, opacity=.015, shadow xshift=-0.40pt, shadow yshift=-0.45pt, shadow scale=1.020,color=black},%
  drop shadow={blend mode=color burn, opacity=.015, shadow xshift=-0.40pt, shadow yshift=-0.45pt, shadow scale=1.015,color=black},%
  drop shadow={blend mode=color burn, opacity=.015, shadow xshift=-0.40pt, shadow yshift=-0.45pt, shadow scale=1.010,color=black},%
  rectangle, rounded corners, ultra thin, draw opacity=1, draw=black!50,  inner sep=2pt, font={\scriptsize\boldmath}, label distance=1mm, text opacity=1,minimum height=.3cm,minimum width=.3cm, fill=white, fill opacity=.85, %
  tikzit category=ZX, tikzit fill=white, tikzit draw=white  %
]
\tikzstyle{bphase}=[
  node on layer= frontlayer, %
  drop shadow={blend mode=color burn, opacity=.03, shadow xshift=-0.40pt, shadow yshift=-0.45pt, shadow scale=1.020,color=black},%
  drop shadow={blend mode=color burn, opacity=.03, shadow xshift=-0.40pt, shadow yshift=-0.45pt, shadow scale=1.015,color=black},%
  drop shadow={blend mode=color burn, opacity=.03, shadow xshift=-0.40pt, shadow yshift=-0.45pt, shadow scale=1.010,color=black},%
  rectangle, rounded corners, ultra thin, draw opacity=1, fill opacity=.7, draw=black!70, inner sep=2pt, font={\scriptsize\boldmath}, label distance=1mm, text opacity=1,minimum height=.3cm,minimum width=.3cm, text=white, fill=black!85, %
  tikzit category=ZX, tikzit fill=black, tikzit draw=white %
]
\tikzstyle{mphase}=[
  node on layer= frontlayer,%
  drop shadow={blend mode=color burn, opacity=.025, shadow xshift=-0.40pt, shadow yshift=-0.45pt, shadow scale=1.020,color=black},%
  drop shadow={blend mode=color burn, opacity=.025, shadow xshift=-0.40pt, shadow yshift=-0.45pt, shadow scale=1.015,color=black},%
  drop shadow={blend mode=color burn, opacity=.025, shadow xshift=-0.40pt, shadow yshift=-0.45pt, shadow scale=1.010,color=black},%
  rectangle, rounded corners, ultra thin, draw opacity=1, draw=black!60, inner sep=2pt, font={\scriptsize\boldmath}, label distance=1mm, fill opacity=.75, text opacity=1,minimum height=.3cm,minimum width=.3cm, fill=black!25,  %
  tikzit category=ZX, tikzit fill=gray, tikzit draw=white %
]
\tikzstyle{lmat}=[shape=signal, signal to=west, signal from=east, fill={zx_grey}, draw=black, minimum height=6pt, inner sep=1pt, font={\scriptsize\boldmath}, tikzit fill=gray, tikzit category=GLA, anchor=center, outer sep=-.1cm, signal pointer angle=\arrowangle,scale=\arrowscaling]
\tikzstyle{rmat}=[shape=signal, signal to=east, signal from=west, fill={zx_grey}, draw=black, minimum height=6pt, inner sep=1pt, font={\scriptsize\boldmath}, tikzit fill=gray, tikzit category=GLA, anchor=center, outer sep=-.1cm, signal pointer angle=\arrowangle,scale=\arrowscaling]
\tikzstyle{dmat}=[shape=signal, signal to=east, signal from=west, fill={zx_grey}, draw=black, minimum height=6pt, inner sep=1pt, font={\scriptsize\boldmath}, tikzit fill=gray, tikzit category=GLA, rotate=270, anchor=center, outer sep=-.1cm, signal pointer angle=\arrowangle,scale=\arrowscaling]
\tikzstyle{umat}=[shape=signal, signal to=east, signal from=west, fill={zx_grey}, draw=black, minimum height=6pt, inner sep=1pt, font={\scriptsize\boldmath}, tikzit fill=gray, tikzit category=GLA, rotate=90, anchor=center, outer sep=-.1cm, signal pointer angle=\arrowangle,scale=\arrowscaling]
\tikzstyle{glmat}=[shape=signal, signal to=west, signal from=east, fill={zx_grey}, draw=black, minimum height=6pt, inner sep=1pt, font={\scriptsize\boldmath}, tikzit fill=gray, tikzit category=GLA, anchor=center, outer sep=-.1cm, signal pointer angle=\arrowangle,scale=\arrowscaling, line width=.8pt, tikzit draw=blue]
\tikzstyle{grmat}=[shape=signal, signal to=east, signal from=west, fill={zx_grey}, draw=black, minimum height=6pt, inner sep=1pt, font={\scriptsize\boldmath}, tikzit fill=gray, tikzit category=GLA, anchor=center, outer sep=-.1cm, signal pointer angle=\arrowangle,scale=\arrowscaling, line width=.8pt, tikzit draw=blue]
\tikzstyle{gdmat}=[shape=signal, signal to=east, signal from=west, fill={zx_grey}, draw=black, minimum height=6pt, inner sep=1pt, font={\scriptsize\boldmath}, tikzit fill=gray, tikzit category=GLA, rotate=270, anchor=center, outer sep=-.1cm, signal pointer angle=\arrowangle,scale=\arrowscaling, line width=.8pt, tikzit draw=blue]
\tikzstyle{gumat}=[shape=signal, signal to=east, signal from=west, fill={zx_grey}, draw=black, minimum height=6pt, inner sep=1pt, font={\scriptsize\boldmath}, tikzit fill=gray, tikzit category=GLA, rotate=90, anchor=center, outer sep=-.1cm, signal pointer angle=\arrowangle,scale=\arrowscaling, line width=.8pt, tikzit draw=blue]
\tikzstyle{box}=[fill=white, draw=black, shape=rectangle, inner sep=2.5pt]
\tikzstyle{wirelable}=[node on layer= labeltextlayer, font={\scriptsize},scale=.9,text=dark_grey, fill=none, inner sep=1pt]
\tikzstyle{gather}=[minimum size=0pt,inner sep=0pt, outer sep=0pt]%
\tikzstyle{divide}=[minimum size=0pt,inner sep=0pt, outer sep=0pt]%
\tikzstyle{multiplexer}=[-, fill={rgb,255: red,179; green,179; blue,179}]

\tikzstyle{gn}=[font={\scriptsize\boldmath}, inner sep=1mm, outer sep=-1.8mm, scale=0.8, tikzit shape=circle, draw=black, fill={zx_green}, tikzit draw=black, tikzit fill=green, tikzit category=ZX, shape=circle]
\tikzstyle{rn}=[font={\scriptsize\boldmath}, inner sep=1mm, outer sep=-1.8mm, scale=0.8, tikzit shape=circle, draw=black, fill={zx_red}, tikzit fill=red, tikzit draw=black, shape=circle, tikzit category=ZX]
\tikzstyle{ggn}=[font={\scriptsize\boldmath}, inner sep=1mm, outer sep=-1.8mm, scale=0.8, tikzit shape=circle, draw=black, fill={zx_green}, tikzit draw=blue, tikzit fill=green, tikzit category=ZX, shape=circle, line width=.8pt]
\tikzstyle{grn}=[font={\scriptsize\boldmath}, inner sep=1mm, outer sep=-1.8mm, scale=0.8, tikzit shape=circle, draw=black, fill={zx_red}, tikzit fill=red, tikzit draw=blue, shape=circle, tikzit category=ZX, line width=.8pt]
\tikzstyle{had}=[fill=yellow, draw=black, shape=rectangle, tikzit category=ZX, tikzit fill=yellow, tikzit draw=black, inner sep=1.5pt, minimum height = 5pt, minimum width = 5pt, font={\scriptsize\boldmath}]
\tikzstyle{ghad}=[fill=yellow, draw=black, shape=rectangle, tikzit category=ZX, tikzit fill=yellow, tikzit draw=blue, inner sep=1.5pt, minimum height = 5pt, minimum width = 5pt, font={\scriptsize\boldmath}, line width=.8pt]

\tikzstyle{gwn}=[style=grn]
\tikzstyle{gbn}=[style=ggn]
\tikzstyle{dmatt}=[style=gdmat]
\tikzstyle{umatt}=[style=gumat]
\tikzstyle{lmatt}=[style=glmat]
\tikzstyle{rmatt}=[style=grmat]
\tikzstyle{scalar}=[style=mphase]

\tikzstyle{gphase}=[rounded rectangle, rounded rectangle arc length=120, fill={zx_green}, inner sep=2pt, font={\tiny\boldmath}, label distance=1mm, fill opacity=.8, text opacity=1, tikzit category=ZX, tikzit fill=green, tikzit draw=green]
\tikzstyle{rphase}=[rounded rectangle, rounded rectangle arc length=120, fill={zx_red}, inner sep=2pt, font={\tiny\boldmath}, label distance=1mm, fill opacity=.6, text opacity=1, tikzit category=ZX, tikzit fill=red, tikzit draw=red]

\tikzstyle{dot}=[thick, fill=black, circle, scale=1, inner sep=.05cm]

\tikzstyle{background}=[-, fill={rgb,255: red,245; green,245; blue,245}, draw=none, tikzit draw={rgb,255: red,128; green,128; blue,128}, on layer= backlayer]
\tikzstyle{white}=[-, fill=white, draw=none, tikzit fill=white]
\tikzstyle{bbox}=[-, fill=white]
\tikzstyle{thick}=[-, line width=.8pt, tikzit draw=blue]
\tikzstyle{very thick}=[-, line width=.8pt, tikzit draw=blue]

\usepackage[
  style=alphabetic,
  sorting=nyt,
  backend=biber,
  maxnames=6,
  maxcitenames=2,
  url=false,
  doi=true,
  isbn=false,
  eprint=false
]{biblatex}

\usepackage{hyperref}
\hypersetup{
  colorlinks=true,
  linkcolor=cyan,
  citecolor=gray
}

\newcommand{\N}{\mathbb{N}}

\newcommand{\R}{\mathbb{R}}
\newcommand{\C}{\mathbb{C}}

\DeclareMathSymbol{\wedge}{\mathbin}{symbols}{"5E}
\DeclareMathSymbol{\vee}{\mathbin}{symbols}{"5F}

\NewDocumentCommand{\sigmalg}{ O{X} }{
  {#1, \sigma_#1}
}

\newcommand{\proofsection}[1]{\medskip\noindent\emph{\(\triangleright\) #1.}}

\newcommand{\btimes}{\mathbin{\hat{\otimes}}}
\newcommand{\wtimes}{\mathbin{\bar{\otimes}}}

\newcommand{\meas}{\mathsf{Measurable}}

\newcommand{\Wstar}{\mathsf{W^*_{CPSU}}}
\NewDocumentCommand{\NCP}{ O{A} O{B} }{%
  \mathsf{W^*_{CP}}(\mathcal{#1},\mathcal{#2})
}
\NewDocumentCommand{\NCB}{ O{A} O{B} }{%
  \mathsf{W^*_{CB}}(\mathcal{#1},\mathcal{#2})
}
\NewDocumentCommand{\NCBSU}{ O{A} O{B} }{%
  \mathsf{W^*_{CBSU}}(\mathcal{#1},\mathcal{#2})
}
\NewDocumentCommand{\CP}{ O{A} O{B} }{%
  \mathsf{{CP}}(\mathcal{#1},\mathcal{#2})
}
\NewDocumentCommand{\CB}{ O{A} O{B} }{%
  \mathsf{{CB}}(\mathcal{#1},\mathcal{#2})
}
\NewDocumentCommand{\CBSU}{ O{A} O{B} }{%
  \mathsf{{CBSU}}(\mathcal{#1},\mathcal{#2})
}
\NewDocumentCommand{\CPSU}{ O{A} O{B} }{%
  \Wstar(\mathcal{#1},\mathcal{#2})
}

\NewDocumentCommand{\monad}{ O{A} O{B} }{%
  Q^{\mathcal{#1}, \mathcal{#2}}
}

\NewDocumentCommand{\evmap}{ O{a} O{b} }{%
  \operatorname{ev}_{{#1}, {#2}}
}
\newtheorem{theorem}{Theorem}
\newtheorem{definition}[theorem]{Definition}
\newtheorem{proposition}[theorem]{Proposition}
\newtheorem{lemma}[theorem]{Lemma}
\newtheorem{corollary}[theorem]{Corollary}

\theoremstyle{definition}
\newtheorem{example}[theorem]{Example}

\declaretheoremstyle[
headfont=\normalfont\bfseries\color{gray},
bodyfont=\normalfont,
notefont=\normalfont\bfseries\color{gray},
notebraces={}{},
headpunct={.},
qed=\qedsymbol,
mdframed={
  linewidth=1.5,
  linecolor=gray,
  hidealllines=true,
  leftline=true,
  skipabove=0,
  innerleftmargin=2mm,
  innerrightmargin=0,
  innertopmargin=0,
  innerbottommargin=.7mm
}
]{line_proof}

\declaretheorem[name=Proof,style=line_proof,numbered=no]{lproof}

\title{\textbf{Composing Quantum Instruments}}

\author[1]{Robert I. Booth}
\author[2]{Dominik Leichtle}
\author[2]{Alex Rice}
\author[2]{Kim Worrall}

\affil[1]{
  University of Oxford, United Kingdom
}
\affil[2]{
  University of Edinburgh, United Kingdom
}
\date{}

\begin{document}

\maketitle

\begin{abstract}
  We study the composition of classically-controlled quantum instruments---the
  natural quantum analogue of Markov kernels. Classically, Markov kernels
  compose by integrating one kernel against another. Defining this composition
  for quantum instruments with continuous outcomes requires an integral of
  quantum channel-valued functions with respect to a quantum instrument. We
  construct this integral in the Heisenberg picture using the Okamura--Ozawa
  normal extension to a von Neumann tensor product. This integral recovers
  the expected finite formula, preserves normal complete positivity and
  subunitality, and provides the multiplication for a monad governing the
  composition of quantum instruments. As an immediate consequence, we identify
  the category of quantum Markov kernels as the Kleisli category of this monad.
\end{abstract}

Quantum measuring processes can be conceptually understood as consisting of
two pieces of data: the probability distributions of classical outcomes given
a quantum state, and the update to the quantum state conditioned on each
outcome. Quantum instruments offer a mathematically elegant formulation
of this viewpoint, decomposing the measuring process into a family of
trace-non-increasing channels for each classical outcome that simultaneously
track both data. A theory of quantum measurement with outcomes in an arbitrary
measure space was formulated by Davies and Lewis \cite{davies:instruments},
who gave a general notion of quantum instruments as quantum channel-valued
measures.

This makes classically-controlled quantum instruments a natural analogue of Markov kernels in settings
where classical and quantum data interact. A classical Markov kernel assigns to
each input a probability measure on possible outputs. A quantum Markov kernel,
in the sense considered here, assigns to each classical input a Davies-Lewis
quantum instrument: the classical input controls a measuring process, whose
outcomes are classical and whose effects on the quantum system are described
by completely positive maps. Equivalently, such kernels may be regarded as
quantum processes acting on both classical and quantum data.  This begs
the immediate question: how do these quantum Markov kernels compose? In particular,
conditioning one quantum Markov kernel on the outcomes of another,
what is the overall resulting operation?

For ordinary Markov kernels, composition is defined by integration: to
compose a kernel \(k : X \times \sigma_Y \to [0,1]\) with a kernel \(h :
Y \times \sigma_Z \to [0,1]\), one integrates the measurable function \(y
\mapsto h(y, E)\) against the measure \(F \mapsto k(x, F)\). This operation
is essentially the multiplication of the Giry monad \cite{giry:monad}. In the
quantum case, the same idea suggests integrating a channel-valued function
against a quantum instrument.  However, this is no longer an ordinary scalar
or even Banach-valued integral. The integrator is itself an instrument,
its values are completely positive maps, and the order of composition matters.

This paper develops the integral needed for this composition. Working in
the Heisenberg picture, we regard quantum channels as normal completely
positive subunital (nCPSU) maps between W*-algebras. A quantum instrument
is then a countably additive measure taking values in nCPSU maps. For
an instrument satisfying the Okamura--Ozawa normal extension property
\cite{okamura_measurement_2015}, the instrument admits a normal completely
positive extension to a W*-tensor product. This extension allows us to define
a noncommutative integral of bounded measurable channel-valued functions
against the instrument.

With this integral in hand, we can construct a quantum analogue of the
(stateful) Giry monad, or \emph{quantum instrument monad}, on the category
of measurable spaces and measurable functions. This monad is parameterised in
the sense of Atkey \cite{atkey:parameterised_monads}, and the Kleisli category of this monad models
quantum Markov kernels.

\paragraph{Structure} In section~\ref{section:motivation}, we motivate
our constructions by briefly overviewing the \emph{finite} quantum
instrument monad, which is a simultaneous generalisation to the quantum
setting of the finite state monad and the finite distribution monad. In
section~\ref{section:algebras}, we summarise some key standard results
from the theory of von Neumann algebras. Section~\ref{section:instruments}
constructs a \emph{noncommutative} integral for an operator-valued function
with respect to an \emph{infinite} quantum instrument, using Okamura-Ozawa's
\emph{normal extension property}. Finally, in section~\ref{section:monad}
we introduce our quantum instrument monad.

\paragraph{Related work}
Independently of the present work, Tobias Fritz has developed a quantum
instrument monad in the Schrödinger picture. The technical details differ---in
particular, the construction of the quantum instrument integral---but the
resulting constructions appear to be closely related. We refer the reader to
\cite{fritz_quantum_2026} for this complementary perspective.

\section{Motivation}
\label{section:motivation}
A quantum measurement process is commonly described by a \emph{quantum
instrument}: a finite family of completely positive maps \(\{\Phi_x\}\)
such that \(\sum_x \Phi_x\) is trace-preserving. The quantity
\(\operatorname{tr}[\Phi_x(\rho)]\) gives the probability of
obtaining outcome \(x\) in state \(\rho\), while the normalized state
\(\Phi_x(\rho)/\operatorname{tr}[\Phi_x(\rho)]\) describes the corresponding
post-measurement state. Thus, the family \(\{\Phi_x\}\) simultaneously
encodes both the measurement statistics and the state update induced by
the measurement.

\begin{example}
  Let \(\{P_i\}\) be an orthogonal family of projectors such that \(\sum_i P_i =
  1\), then the quantum instrument associated with this measurement is given
  by the family \(\Phi_i(\rho) = P_i \rho P_i\).
\end{example}

For infinite-dimensional quantum systems, it will be convenient to use the
Heisenberg picture. Thus a measurement process from a system with Hilbert
space \(\mathcal H\) to one with Hilbert space \(\mathcal K\), with outcomes
in a finite or countable set \(X\), is described by a family of completely
positive maps
\begin{equation}
  \Phi_x:\mathcal B(\mathcal K)\to\mathcal B(\mathcal H)
\end{equation}
such that \(\sum_x\Phi_x\) is unital. Given an initial state \(\omega\) on
\(\mathcal B(\mathcal H)\), the probability of obtaining outcome \(x\) is
\begin{equation}
  p_x := \omega(\Phi_x(1_{\mathcal K})).
\end{equation}
Note that the positivity of $\Phi_x$ guarantees $p_x \geq 0$ while the normalisation (unitality) of $\sum_x \Phi_x$ implies that $\sum_x p_x = 1$.
Conditioned on this outcome, the resulting state on \(\mathcal B(\mathcal K)\)
is
\begin{equation}
  \omega_x(A)
  =
  \frac{\omega(\Phi_x(A))}
       {\omega(\Phi_x(1_{\mathcal K}))},
  \qquad A\in\mathcal B(\mathcal K),
\end{equation}
provided the denominator is nonzero. 

The preceding discussion suggests that quantum instruments possess a rich
compositional structure. A measurement process may be followed by a second
measurement, and the choice of the second measurement may itself depend on the
outcome of the first. Consequently, one would like a mathematical formalism
in which measurement procedures can be assembled into more complicated
measurement procedures.

Two distinct mechanisms are at work. First, a measurement transforms the
quantum state, so that the output state of one instrument becomes the input
state of the next. Second, a measurement produces outcomes with specified
probabilities, introducing a form of probabilistic branching.

These two phenomena are familiar from other contexts. State evolution is
captured categorically by the state monad and its parameterised variants
\cite{moggi:notions,atkey:parameterised_monads}, while probabilistic branching is captured by probability monads
such as the finite distribution monad \cite{fritz:distribution_monad}. Quantum instruments combine
both features in a single object: they describe probabilistic processes which
also transform an underlying quantum state.

Moreover, both aspects participate in the composition of measurement
processes. States evolve sequentially through successive instruments, while
measurement outcomes determine how the process branches and which subsequent
instrument is applied. The finite-dimensional instrument construction
considered below provides a mathematical framework in which these two
forms of composition coexist. The resulting structure carries the form of a
parameterised monad, simultaneously extending the parameterised state monad
and the finite distribution monad.  This finite-dimensional setting serves
as a useful toy model for the more general quantum instrument monad developed
in the remainder of the paper.

We give the formal definitions of monads and parameterised monads in
appendix~\ref{appendix:monads}, and focus here on motivating the monads we
are interested in for modelling quantum Markov kernels.

\subsection{The parameterised state monad}

Suppose a physical system has state space \(S\). A \emph{perturbative observation}
of the system naturally modelled by a function
\begin{equation}
  S \longrightarrow X \times S.
\end{equation}
which returns an observable outcome in some set \(X\), while possible altering
the state of the system. Equivalently, the collection of all perturbative
observations of \(S\) with outcomes in \(X\) is
\begin{equation}
  M^S X \coloneqq \mathsf{Set}(S,X\times S).
\end{equation}
In general, when performing a sequence of observations of the system \(S\), one
might vary the choice of observations as a function of the outcome of previous
observations. This endows the assignment \(X \mapsto M^S X\) with a non-trivial
compositional structure which is naturally captured by the \emph{state monad}.

In many situations, however, the type of the state itself need not remain
fixed. An observation may transform states of one type into states of
another, as is common for quantum channels. This leads naturally to Atkey's
\emph{parameterised state monad} \cite{atkey:parameterised_monads}, defined by
\begin{equation}
  M^{S,T}X \coloneqq \mathsf{Set}(S,X\times T).
\end{equation}
Here an observation begins with a state in \(S\), produces an outcome in
\(X\), and leaves the system in a state belonging to \(T\). As before,
observations can be composed whenever the output state space of one agrees
with the input state space of the next. This composition equips the family
\(\{M^{S,T}\}\) with the structure of a parameterised monad. Operationally,
the multiplication corresponds to performing an observation whose outcome
determines a subsequent observation.

The parameterised state monad provides a natural model of stateful physical
processes whose state space may change. This is precisely the situation
encountered for quantum instruments, where measurements may transform states
between different Hilbert spaces.

\subsection{The finite distribution monad}

Just as the state monad captures stateful processes, probability
monads capture probabilistic branching. The simplest example is the finite
distribution monad. Given a set \(X\), let \(\mathcal{D}(X)\) denote the set
of finitely supported probability distributions on \(X\). A function
\begin{equation}
  f : X \to \mathcal{D}(Y)
\end{equation}
may be interpreted as a probabilistic process: given an input \(x\),
the process returns an output distributed according to the probability
distribution \(f(x)\).

The unit of the monad sends an element to the corresponding Dirac distribution,
\begin{equation}
   \eta_X(x)=\delta_x,
\end{equation}
while the multiplication
\begin{equation}
   \begin{aligned}
     \mu_X : \mathcal{D}(\mathcal{D}(X)) &\longrightarrow \mathcal{D}(X) \\
     p &\longmapsto \sum_{d \in \mathcal{D}(X)} p_d \cdot d
   \end{aligned}
\end{equation}
averages a probability distribution over probability distributions into
a single probability distribution. Operationally, this corresponds to
performing a probabilistic choice whose outcomes are themselves probabilistic
processes.

This monad provides a categorical model of probabilistic branching. In
particular, sequential composition of probabilistic processes is encoded
by the monad multiplication, exactly as sequential composition of stateful
computations is encoded by the state monad.

The finite distribution monad admits a natural extension from sets to
measurable spaces called the \emph{Giry monad} \cite{giry:monad}, in which
finitely supported distributions are replaced by probability measures. The
Giry monad plays a central role in categorical probability theory and will
provide one of the principal ingredients in our construction of the quantum
instrument monad.

\subsection{The finite quantum instrument monad}

The preceding discussion suggests that quantum instruments should admit a
parameterised monadic structure. We now describe the finite version of this
construction.

Like the parameterised state monad, the underlying state space is allowed to
vary. In the quantum setting the state spaces are Hilbert spaces, and the state
transformations are completely positive maps. Like the finite distribution
monad, measurements branch over a set of classical outcomes. Accordingly,
for Hilbert spaces \(\mathcal{H,K}\) and an outcome set \(X\), let
\(Q^\mathcal{H,K} X\) be the set of finitely-supported quantum instruments, that is
\(\{\Phi_x : \mathcal{B}(\mathcal{K}) \to \mathcal{B}(\mathcal{H})\}_{x
\in X}\), such that \(\sum_{x \in X} \Phi_x\) is unital.

The units are analogous to the Dirac distribution from the finite distribution
monad. For any Hilbert space \(\mathcal{H}\) and \(x \in X\), define a
\emph{Dirac instrument} as
\begin{equation}
  \left(\delta_x^\mathcal{H}\right)_y
  =
  \begin{cases}
   1_\mathcal{H} \text{ if } x = y; \\
   0 \text{ otherwise.}
  \end{cases}
\end{equation}
where \(1_\mathcal{H}\) is the identity channel. The units of the monad are
given by \(\eta_{X}^\mathcal{H}(x) = \delta_x^\mathcal{H}\).

An element of \(\monad[H][J] \monad[J][K] X\) may be viewed as a
two-stage measurement procedure: one first performs a measurement on
\(\mathcal{H}\), which transforms the quantum system to the Hilbert space
\(\mathcal{J}\), and whose classical outcomes also determine a second
instrument on \(\mathcal{J}\). The multiplication simply regards this
adaptive two-stage procedure as a single instrument from \(\mathcal{H}\)
to \(\mathcal{K}\). Explicitly,
\begin{equation}
  \left(\mu_X^{\mathcal H,\mathcal J,\mathcal K}(p)\right)_x
  =
  \sum_{q \in Q^{\mathcal J,\mathcal K}X}
  p_q \circ q_x.
\end{equation}
Here \(p_q: \mathcal{B}(\mathcal J) \to \mathcal{B}(\mathcal H)\) is
the operation associated with the outer outcome \(q\), while \(q_x:
\mathcal{B}(\mathcal K) \to \mathcal{B}(\mathcal J)\) is the operation
associated with outcome \(x\) of the inner instrument \(q\).
Note that only finitely many of the summands can be non-zero.

This combines the two constructions discussed above. As in the finite
distribution monad, the outer measurement averages over the possible
intermediate outcomes. As in the parameterised state monad, the intermediate
quantum state produced by the first instrument is passed to the second.
\begin{theorem}
  The constructions above endow the family \(\monad[H][K]\) with the structure
  of a parameterised monad.
\end{theorem}
Rather than verifying the axioms directly in this finite setting, we defer
the proof to the general measurable construction developed below, of which
the present monad is a special case.

\section{Elements of W*-algebras and quantum channels}
\label{section:algebras}
\begin{definition}
  A \emph{W*-algebra} is a Banach *-algebra \(\mathcal{A}\) which is
  the topological dual of a Banach algebra \(\mathcal{A}_*\) called
  its \emph{predual}. The weak-* topology induced on \(\mathcal{A}\)
  by the functionals in \(\mathcal{A}_* \subseteq (\mathcal{A}_*)^{**} =
  \mathcal{A}^*\) is called the \emph{ultraweak} topology.
\end{definition}

W*-algebras come with sufficient structure to define a notion of positivity, equipping them with a natural order structure. In the following, we review only some of their properties that are important for our purposes. For a more detailed discussion, we refer to \cite{sakai_c-algebras_2012}.

\begin{definition}
  Let \(\mathcal{A},\mathcal{B}\) be W*-algebras, then a linear map \(\Phi : \mathcal{A} \to \mathcal{B}\) is
  \begin{itemize}
    \item \emph{normal} if it is ultraweakly continuous;
    \item \emph{unital} if \(\Phi(1_\mathcal{A}) = 1_\mathcal{B}\);
    \item \emph{subunital} if \(\Phi(1_\mathcal{A}) \leqslant 1_\mathcal{B}\);
    \item \emph{positive} if \(\Phi(a) \geqslant 0\) for every \(a \geqslant 0\).
  \end{itemize}
\end{definition}

\begin{proposition}[\cite{cho:semantics}]\label{prop:cho_normal_monotone_net}
  Let \(\Phi : \mathcal{A} \to \mathcal{B}\) be positive. Then \(\Phi\) is
  normal if and only if it preserves the supremum of norm-bounded monotone
  nets in \(\mathcal{A}\).
\end{proposition}

Consider W*-algebras \(\mathcal{A,B}\), and a linear map \(\Phi : \mathcal{A}
\to \mathcal{B}\). The \emph{matrix amplifications} of \(\Phi\) are the linear
maps defined for each \(n \in \N\) by:
\begin{equation}
  \begin{aligned}
    \Phi^{(n)} : M_n(\mathcal{A}) &\longrightarrow M_n(\mathcal{B}) \\
    [a_{jk}] &\longmapsto [\Phi(a_{jk})].
  \end{aligned}
\end{equation}

\begin{definition}
  We say that \(\Phi\) is \emph{completely bounded} if each \(\Phi^{(n)}\)
  is bounded and \emph{completely positive} if each \(\Phi^{(n)}\) is positive.
\end{definition}

\begin{proposition}[\cite{cho:semantics}]
  There is a symmetric monoidal category \(\Wstar\) where
  \begin{itemize}
    \item \emph{objects} are W*-algebras;
    \item \emph{arrows} are normal completely positive subunital linear maps;
    \item the \emph{monoidal structure} is given by the spatial von Neumann
    tensor product, or W*-tensor, $\bar{\otimes}$ (see \cite[Definition
    1.22.10]{sakai_c-algebras_2012} for a full definition).
    \end{itemize}
\end{proposition}

We impose throughout that all W*-algebras have a separable predual. The key
consequence of this assumption is lemma~\ref{lemma:instrument_domination},
which informally says that all quantum instruments can be thought of as
densities with respect to a real measure.

\section{The integral theory of quantum instruments}
\label{section:instruments}
Following the discussion of the discrete-outcome case in section~\ref{section:motivation}, we now move towards quantum instruments with continuous outcomes by revisiting Davies' and Lewis' definition of quantum instruments on measurable spaces which generalise both finite-outcome quantum instruments and classical distributions in the form of probability measures.

\begin{definition}[Quantum instrument \cite{davies:instruments}]
Let \(\mathcal{A,B}\) be W*-algebras and \((X, \sigma_X)\) a measurable
space, then a \emph{quantum instrument} \(\mathcal{A} \to_X \mathcal{B}\)
is a mapping \(q : \sigma_X \to \CPSU\) which is
\begin{itemize}
  \item \emph{strict}: \(q(\varnothing) = 0_{\mathcal{A,B}}\);
  \item \emph{normalised}: \(q(X)(1_\mathcal{A}) = 1_\mathcal{B}\);
  \item \emph{ultraweakly countably additive}: for any countable collection \(\{E_j\}
  \subseteq \sigma_X\) of disjoint sets, \(a \in \mathcal{A}\), and \(\phi
  \in \mathcal{B}_*\), \(\langle \phi, q(\bigcup_j E_j)(a) \rangle = \sum_j
  \langle \phi, q(E_j)(a) \rangle\).
\end{itemize}
\end{definition}

In particular, for any quantum instrument \(q : \mathcal{A} \to_X \mathcal{B}\) and
any (respectively, positive unital) \(\phi \in \mathcal{B}_*\),
the mapping \(q_\phi : \sigma_X \to \R_+, \; E \mapsto \langle \phi, q(E)(1_\mathcal{A})
\rangle\) defines a complex (respectively, probability) measure on \(X\).

\begin{lemma}
  \label{lemma:instrument_domination}
  For any quantum instrument \(q : \mathcal{A} \to_X \mathcal{B}\), there is
  a finite real measure \(\nu\) on \(X\) which dominates \(q\). Explicitly,
  \(q(E) = 0_{\mathcal{A,B}}\) whenever \(\nu(E) = 0\).
\end{lemma}
\begin{lproof}
  Since \(\mathcal B_*\) is separable (by assumption), there is a faithful
  positive state \(s \in \mathcal{B}_*\). Define a finite real measure
  \(\nu \coloneqq E \mapsto \langle s, q(E)(1_\mathcal{A}) \rangle\), then
  if \(\nu(E) = \langle s, q(E)(1) \rangle = 0\), we must have \(q(E)(1) =
  0\), which since \(q(E)\) is completely positive implies \(q(E) = 0\).
\end{lproof}

Since quantum instruments are in effect CPSU-valued measures, we would like
to define a \emph{noncommutative} integral of CPSU-valued functions \(\Phi\)
w.r.t. a quantum instrument \(q\)
\begin{equation}
  \int_X q(\dd{x}) \circ \Phi(x),
\end{equation}
which extends the case where \(q\) takes only finitely many values and the
integral reduces to a straightforward sum. This integral suggests the existence
of ``left'' and ``right'' orderings, or pre- and post-composition by \(\Phi\). We
focus on the pre-composition integral since, as we shall see in the sequel,
this is the ordering that corresponds to the physically relevant composition
of quantum stochastic processes. In particular, one can already see in the
finite case that only the pre-composition integral guarantees subunitality of
the resulting CP map.

We construct this integral via the well-known connection between
operator-valued measurable functions, measurable fields of operators,
and tensor products. Many of these statements are either known, expected
or folklore, but we provide proofs for those statements we could not find
verbatim in the literature for the sake of completeness.

\begin{proposition}
  \label{proposition:measurable_tensor}
  Let \((X,\sigma_X)\) be a measurable space,
  \(\nu\) a \(\sigma\)-finite measure on \(X\),
  \(\mathcal{A}\) a W*-algebra,
  and \(\Phi : X \to \mathcal{A}\) a bounded ultraweakly measurable map.
  Then, \(\Phi\) determines a
  unique element \([\Phi]_\nu \in L^\infty(\nu)\,\bar{\otimes}\,\mathcal{A}\) characterised by
  \[
    \left\langle f \otimes \phi, [\Phi]_\nu\right\rangle
    =
    \int_X f(x) \cdot\langle \phi,\Phi(x)\rangle \ \nu(\dd{x}),
    \qq{where}
    f\in L^1(\nu),\ \phi\in \mathcal A_*.
  \]
\end{proposition}

Note that Proposition~\ref{proposition:measurable_tensor} holds in the slightly more general case in which $\Phi$ is only essentially bounded, up to null-sets of $\nu$, rather than bounded everywhere.
Since we are interested in the case in which the integral of $\Phi$ can be defined along \emph{any} quantum instrument, we restrict to the latter case in the following.

\begin{lproof}
  For each \(\phi \in \mathcal{A}_*\), the map \(X \to \C : x \mapsto \langle
  \phi, \Phi(x) \rangle\) is bounded and measurable. It therefore determines
  a bounded bilinear map
  \begin{equation}
    \begin{aligned}
      L^1(\nu) \times \mathcal{A}_* &\longrightarrow \C \\
      (f,\phi) &\longmapsto \int_X f(x) \cdot \langle \phi, \Phi(x) \rangle \  \nu(\dd{x}),
    \end{aligned}
  \end{equation}
  which, by the universal property of the projective tensor product
  \cite[Theorem 2.7.4]{helemskii_lectures_2006}, extends uniquely to a
  bounded linear map \(L^1(\nu) \btimes \mathcal{A}_* \to \C\).  Recalling
  the identifications \(L^\infty(\nu) \,\bar{\otimes}\, \mathcal{A} \cong
  L^1(\nu,\mathcal{A}_*)^*\) \cite[Theorem 1.22.13]{sakai_c-algebras_2012} and
  \(L^1(\nu,\mathcal{A}_*) \cong L^1(\nu) \,\hat{\otimes}\, \mathcal{A}_*\)
  \cite[Theorem 1.22.12]{sakai_c-algebras_2012} this yields an element
  \([\Phi]_\nu \in L^\infty(\nu) \,\bar{\otimes}\, \mathcal{A}\) modulo
  null-sets of \(\nu\), as claimed.
\end{lproof}

As a result, any reasonable integral with respect to a quantum instrument \(q
: \mathcal{A} \to_X \mathcal{B}\) should be given by a map \(\overline{q}
: L^\infty(\nu) \wtimes \mathcal{A} \to \mathcal{B}\).  Simple functions
\(s \coloneqq \sum_k a_k \cdot 1_{E_k}\), where \(a_k \in \mathcal{A}\)
and \(1_{E_k}\) is the characteristic function of the measurable set \(E_k
\in \sigma_X\), get mapped to \([s]_\nu = \sum_k 1_{E_k} \otimes a_k\). In
other words, approximating arbitrary integrands by measurable functions in
the ultraweak sense, we should expect this function \(\overline{q}\) to be
a normal completely positive unital linear map.

\textcite{okamura_measurement_2015} show that every quantum
instrument \(\mathcal{A} \to_X \mathcal{B}\) is equivalent to a unital
completely positive map \(L^\infty(\nu)\,\otimes_\mathrm{bin}\, \mathcal{A}
\to \mathcal{B}\), where \(\otimes_\mathrm{bin}\) is the \emph{C*-algebraic
binormal tensor}. In order to respect the measurable i.e. W*-algebraic
structure we are interested in, we would like this linear map to extend to
a \emph{normal} map \(L^\infty(\nu) \wtimes \mathcal{A} \to \mathcal{B}\).

\begin{definition}[Normal extension property \cite{okamura_measurement_2015}]
  A quantum instrument \(q : \mathcal{A} \to_X \mathcal{B}\) has the
  \emph{normal extension property} (NEP) if there exists a finite measure
  \(\nu_\rho\) on \(X\) and a normal completely positive unital map
  \(\overline{q} : L^\infty(\nu_\rho) \wtimes \mathcal{A} \to \mathcal{B}\)
  such that \(q(E)(a) = \overline{q}(1_E \otimes a)\) for all \(E \in
  \sigma_X\) and \(a \in \mathcal{A}\).
\end{definition}

In order to simplify notation, for any bounded measurable \(\Phi : X \to
\mathcal{A}\), we write \([\Phi]_q \coloneqq [\Phi]_{\nu_q}\) the
measurable field obtained from proposition~\ref{proposition:measurable_tensor}.
If \(q\) has the NEP, we can straightforwardly define an integral of \(\Phi\)
with respect to \(q\) as
\begin{equation}
  \int_X q(\dd{x}) \circ \Phi(x) \coloneqq \overline{q}([\Phi]_q).
\end{equation}
Unfortunately, according to Okamura and Ozawa, ``usually it is not easy to
check whether a given CP instrument has the NEP or not''. They prove that
quantum instruments whose codomain is an \emph{atomic} W*-algebra always have
the NEP. However, outside this case, simple counterexamples already exist,
such as sharp measurements on \(L^\infty([0,1])\).

\begin{definition}
  A map \(\Phi : X \to \CPSU[A][B]\) is \emph{measurable} if for all \(\phi \in
  \mathcal{B}_*\) and \(a \in \mathcal{A}\), the mapping \(x \mapsto \langle \phi,
  \Phi(x)(a) \rangle\) is Borel measurable.
\end{definition}

In particular, for any bounded measurable \(\Phi : X \to \CPSU\), and each
\(a \in \mathcal{A}\), the map \(\Phi(-)(a) : X \to \mathcal{B} : x \mapsto
\Phi(x)(a)\) is bounded and ultraweakly measurable, so we can define an
integral along an NEP quantum instrument $q : \mathcal{B} \to_X \mathcal{C}$ pointwise as
\begin{equation}
  \left(\int_X q(\dd{x}) \circ \Phi \right)(a) \coloneqq \overline{q}([\Phi(-)(a)]_q).
\end{equation}

\begin{proposition}
  \label{proposition:integral_CPSU}
  For any bounded measurable map \(\Phi : X \to \CPSU[A][B]\) and NEP quantum
  instrument \(q : \mathcal{B} \to_X \mathcal{C}\), \(\int_X q(\dd{x})
  \circ \Phi\) is a normal completely positive subunital map \(\mathcal{A}
  \to \mathcal{C}\).
\end{proposition}
\begin{lproof}
  By assumption, we have an integral \(\overline{q} :
  L^\infty(\nu_q) \wtimes \mathcal{B} \to \mathcal{C}\) and by
  proposition~\ref{proposition:measurable_tensor} a map
  \begin{equation}
    \begin{aligned}
      \overline{\Phi} : \mathcal{A} &\longrightarrow L^\infty(\nu_q) \wtimes \mathcal{B} \\
      a &\longmapsto [\Phi(-)(a)]_q
    \end{aligned}
  \end{equation}
  which is uniquely defined by the property that
  \begin{equation}
    \left\langle f \otimes \phi, \overline{\Phi}(a) \right\rangle
    =
    \int_X f(x) \cdot\langle \phi,\Phi(x)(a)\rangle \ \nu_q(\dd{x}),
    \qq{where}
    f\in L^1(\nu_q),\ \phi\in \mathcal B_*,
  \end{equation}
  so \(\overline{\Phi}\) is linear and bounded.

  \proofsection{\(\overline{\Phi}\) is subunital} 
  For any positive \(f \in L^1(\nu_q)\) and positive \(\phi \in \mathcal{B}_*\),
  \begin{align}
    \left\langle f \otimes \phi, \overline{\Phi}(1_\mathcal{A}) \right\rangle
    &= \int_X f(x) \cdot\langle \phi,\Phi(x)(1_\mathcal{A})\rangle \ \nu_q(\dd{x}), \\
    &\leqslant \left\langle \phi, 1_\mathcal{B} \right\rangle \cdot \int_X f(x) \ \nu_q(\dd{x}) \\
    &= \left\langle f \otimes \phi, 1_X \otimes 1_\mathcal{B} \right\rangle,
  \end{align}
  so \(\overline{\Phi}\) is subunital.

  \proofsection{\(\overline{\Phi}\) is completely positive}
  Consider the matrix amplification \(\overline{\Phi}^{(n)}\), which is uniquely
  defined by \(\overline{\Phi}^{(n)}([a_{jk}]) = [[\Phi(-)(a)]_q]_{jk} \),
  so that in particular, for any \(f \in L^1(\nu_q)\) and \([\phi_{jk}] \in
  M_n(\mathcal{B_*})\),
  \begin{align}
    \left\langle f \otimes [\phi_{jk}], \overline{\Phi}^{(n)}([a_{uv}]) \right\rangle
    &= \left\langle f \otimes [\phi_{jk}],  [[\Phi(-)(a)]_q]_{jk}) \right\rangle \\
    &= \int_X f(x) \cdot\langle [\phi_{jk}], [\Phi(x)(a)]_{jk}) \rangle \ \nu_q(\dd{x}) \\
    &= \int_X f(x) \cdot\langle [\phi_{jk}], \Phi(x)^{(n)}([a_{uv}]) \rangle \ \nu_q(\dd{x}).
  \end{align}

  Recalling the identification \(M_n(L^\infty(\nu_q) \wtimes
  \mathcal{B})_* \cong L^1(\nu_q) \btimes M_n(\mathcal{B}_*)\), for any
  positive \(f \in L^1(\nu_q)\), positive \([a_{uv}] \in M_n(\mathcal{A})\)
  and positive \([\phi_{jk}] \in M_n(\mathcal{B_*})\), 
  \begin{align}
    \left\langle f \otimes [\phi_{jk}], \overline{\Phi}^{(n)}([a_{uv}]) \right\rangle
    &= \int_X f(x) \cdot\langle [\phi_{jk}], \Phi(x)^{(n)}([a_{uv}]) \rangle \ \nu_q(\dd{x}),
  \end{align}
  which is positive since \(\Phi(x)^{(n)} : M_n(\mathcal{A}) \to
  M_n(\mathcal{B})\) is positive by assumption.

  \proofsection{\(\overline{\Phi}\) is normal}
  Consider a positive monotone net \(a_k \to a\) in \(\mathcal{A}\),
  then by the monotone convergence theorem, for any positive \(f \in L^1(\nu_q)\)
  and \(\phi \in \mathcal{B}_*\),
  \begin{align}
    \left\langle f \otimes \phi, \overline{\Phi}(a_k) \right\rangle
    &= \int_X f(x) \cdot \langle \phi, \Phi(x)(a_k) \rangle \ \nu_q(\dd{x}) \\
    &\to \int_X f(x) \cdot \langle \phi, \Phi(x)(a) \rangle \ \nu_q(\dd{x}) \\
    &= \left\langle f \otimes \phi, \overline{\Phi}(a) \right\rangle,
  \end{align}
  and this extends to evaluation with respect to any element of \(L^1(\nu_q)
  \btimes \mathcal{B}_*\) by norm density of algebraic tensors. Therefore,
  \(\overline{\Phi}(a_k) \to \overline{\Phi}(a)\) and \(\overline{\Phi}\) is normal
  by Proposition~\ref{prop:cho_normal_monotone_net}.

  \medskip
  Finally, as the composition of normal completely positive subunital maps,
  \(\int_X q(\dd{x}) \circ \Phi(x) = \overline{q} \circ \overline{\Phi}\)
  is itself a normal completely positive subunital map.
\end{lproof}

\section{The quantum instrument monad}
\label{section:monad}
\subsection{The parameterised functor}

The \emph{quantum instrument monad} is defined on the \(\Wstar\)-indexed
functor \(\monad\) on \(\mathsf{Measurable}\), which maps \((X,\sigma_X) \in
\mathsf{Measurable}\) to the set of all NEP quantum instruments \(\mathcal{A}
\to_X \mathcal{B}\) over \((X, \sigma_X)\). We equip this set with the coarsest
\(\sigma\)-algebra that makes the evaluation mappings
\begin{equation}
  \begin{aligned}
    \operatorname{ev}_{U,a,b} : \monad(X,\sigma_X) &\longrightarrow \C \\
		q &\longmapsto \langle a, q(U)(b) \rangle
  \end{aligned}
\end{equation}
Borel-measurable for each \(U \in \sigma_X\), \(a \in \mathcal{A}_*\), and \(b \in
\mathcal{B}\).

Its action on morphisms is given in the expected way by \emph{pushforward}: for
any \(f \in \mathsf{Measurable}(X,Y)\) we define a map \(\monad(f) : \monad(X)
\to \monad(Y)\), given for any quantum instrument \(q \in \monad(X)\)
and \(E \in \sigma_Y\) by \(\monad(f)(q)(E) \coloneqq q(f^{-1}(E))\). Since
\begin{equation}
  \operatorname{ev}_{B,a,b} \circ \monad(f)(q) = \langle a, q(f^{-1}(B))(b) \rangle
  = \operatorname{ev}_{f^{-1}(B),a,b} (q)
\end{equation}
for any \(B \in \sigma_Y\), \(a \in \mathcal{A}_*\), \(b \in \mathcal{B}\)
and \(q \in \monad(X,\sigma_X)\), \(\monad(f)\) is measurable.

\subsection{The monad structure}

Following the discussion of section~\ref{section:motivation}, we want to show
that \(\monad\) admits a parameterised monad structure. By analogy with the
Giry monad, for any \(\mathcal{A} \in \Wstar\) and \(x \in X\), define a
\emph{Dirac instrument} by
\begin{equation}
  \begin{aligned}
    \delta_x^\mathcal{A} : \sigma_X &\longrightarrow \CPSU[\mathcal{A}][\mathcal{A}] \\
    U &\longmapsto \begin{cases} 1_\mathcal{A} \text{ if } x \in U; \\ 0 \text{ otherwise.} \end{cases}
  \end{aligned}
\end{equation}

The units of the monad are given by \(\eta_{X}^\mathcal{A} :
X \to \monad[A][A](X) : x \mapsto \delta_x^\mathcal{A}\).

\begin{lemma}
  \label{lemma:dirac_NEP}
  For any \(x \in X\), \(\eta_X^\mathcal{A}(x)
  = \delta_x^\mathcal{A}\) has the NEP. Explicitly, for any
  bounded ultraweakly measurable map \(\Phi : X \to \mathcal{A}\),
  \(\overline{\delta_x^\mathcal{A}}([\Phi]_{\delta_x^\mathcal{A}}) = \Phi(x)\).
\end{lemma}
\begin{lproof}
  Denote \(\delta_x\) the Dirac measure at \(x \in X\), then Dirac instruments
  have the NEP, with \(\overline{\delta_x^\mathcal{A}} : L^\infty(\delta_x)
  \wtimes \mathcal{A} \to \mathcal{A}\) given on algebraic tensors by
  \(\overline{\delta_x^\mathcal{A}}(f \otimes a) = f(x) \cdot a\).

  It's straightforward that this mapping is ultraweakly continuous and
  completely positive on algebraic tensors, hence extends uniquely to a normal
  CPU map as claimed.
\end{lproof}

\begin{lemma}
  \(\eta_{X}^\mathcal{A}\) is measurable.
\end{lemma}
\begin{lproof}
  We have \(\operatorname{ev}_{U,a,b} \circ \eta_{X}^\mathcal{A}(x) =
  \operatorname{ev}_{U,a,b}(\delta_x) = \langle a, \delta_x(U) (b) \rangle =
  \langle a, b \rangle \cdot 1_U(x)\) which is measurable since indicator
  functions of measurable sets are always measurable.
\end{lproof}

\begin{lemma}
  The family \((\eta_{X}^\mathcal{A})_X\) describes a natural transformation
  \(1_X \mapsto \monad[A][A]\).
\end{lemma}
\begin{lproof}
  This is entirely analogous to the Giry case.
\end{lproof}

By analogy with the finite case and probability monads, we define the monad
multiplication as the integral
\begin{equation}
  \label{eq:kleisli_extension}
  \begin{aligned}
    \mu^{\mathcal{A,B,C}}_X : \monad[B][C] \monad X &\longrightarrow \monad[A][C] X, \\
    p &\longmapsto \left(E \mapsto \int_{\monad(X)} p(\dd{q}) \circ q(E) \right).
  \end{aligned}
\end{equation}
More explicitly, \(\mu^{\mathcal{A,B,C}}_X(p)(E)(a) = \tilde{p}([q \mapsto
q(E)(a)]_p)\).

\begin{lemma}
  \label{lemma:multiplication_instrument}
  For any \(p \in \monad[B][C] \monad X\), \(\mu^{\mathcal{A,B,C}}_X(p)\)
  is a quantum instrument \(\mathcal{A} \to_X \mathcal{C}\).
\end{lemma}
\begin{lproof}
  \proofsection{\(q \mapsto q(E)(a)\) is bounded}
  For any \(q \in \monad X\), \(E \in \sigma_X\) and \(a \in \mathcal{A}\),
  we have that \(q(E)(a) \leqslant \norm{a} \cdot q(E)(1_\mathcal{A}) \leqslant
  \norm{a} 1_\mathcal{B}\), so that \(\norm{q(E)(a)} \leqslant \norm{a}\).

  \proofsection{\(q \mapsto q(E)(a)\) is ultraweakly measurable}
  For any \(E \in \sigma_X\), \(a \in \mathcal{A}\), and \(b \in
  \mathcal{B}_*\), the mapping \(\monad X \to \C : q \mapsto \langle b,
  q(E)(a) \rangle\) is measurable, by definition of the measurable structure
  on \(\monad X\), so that the mapping \(\monad X \to \mathcal{B} : q \mapsto
  q(E)(a)\) is also measurable.

  \medskip
  Therefore, by proposition~\ref{proposition:integral_CPSU}, the integral
  \begin{equation}
    \mu^{\mathcal{A,B,C}}_X(p)(E)(a) \coloneqq \int_{\monad(X)} p(\dd{q}) \circ q(E)(a) = \overline{p}([q \mapsto q(E)(a)]_p)
  \end{equation}
  is well-defined, and \(\mu^{\mathcal{A,B,C}}_X(p)(E)\) is normal, subunital
  and completely-positive.

  \medskip
  \proofsection{Countable additivity and strictness}
  Let \((E_k)_{k \in \mathbb{N}} \subseteq \sigma_Y\) be pairwise
  disjoint and put \(E = \bigcup_k E_k\), then since any \(q \in \monad
  X\) is subadditivite, \(q(E) = \sum_k q(E_k)\). Then by linearity of
  \(\overline{p}\) and of the mapping \(\Phi \to [\Phi]_p\),
  \begin{align}
    \int_{\monad{X}} p(\dd q) \circ q(E)
    &= \overline{p}([q \mapsto q(E)]_p) \\
    &= \overline{p}([q \mapsto \sum_k q(E_k)]_p) \\
    &=  \sum_k \overline{p}([q \mapsto q(E_k)]_p) \\
    &= \sum_k \int_{\monad{X}} p(\dd q) \circ q(E_k),
  \end{align}
  with ultraweak convergence of the sum. Since for all \(q \in \monad{X}\) we have
  \(q(\varnothing) = 0\), we get \(\int_{\monad{X}} p(\dd q) \circ q(\varnothing)
  = 0\).

  \medskip
  \proofsection{Normalisation}
  This follows almost immediately since
  \begin{align}
    \int_{\monad{X}} p(\dd{q}) \circ q(X)(1_\mathcal{A})
    = \overline{p}([q \mapsto q(X)(1_\mathcal{A})]_p)
    = \overline{p}([q \mapsto 1_\mathcal{B}]_p)
    = \overline{p}(1_{\monad{X}} \otimes 1_\mathcal{B}) = 1_\mathcal{C}.
  \end{align}

  \medskip

  Altogether, we have shown that the assignment
  \begin{equation}
    \sigma_Y \longrightarrow \CPSU[A][C] :
    E \longmapsto  \int_{\monad{X}} \mu(\dd q) \circ q(E)
  \end{equation}
  yields a strict, normalised, countably additive map \(\sigma_Y \to
  \CPSU[A][C]\).  Hence it is a quantum instrument \(\mathcal{A} \to
  \mathcal{C}\) on \((Y,\sigma_Y)\), as claimed.
\end{lproof}

\begin{lemma}
  \label{lemma:multiplication_NEP}
  For any \(p \in \monad[B][C] \monad X\), \(\mu^{\mathcal{A,B,C}}_X(p)\)
  has the NEP. Explicitly, for any bounded ultraweakly measurable \(\Phi :
  \monad X \to \mathcal{B}\),
  \begin{equation*}
    \overline{\mu^{\mathcal A,\mathcal B,\mathcal C}_X(p)}([\Phi]_{\mu(p)})
    =
    \overline p\bigl([q\mapsto \overline q([\Phi]_q)]_p\bigr).
  \end{equation*}
  where \([\Phi]_{\mu(p)}\) is shorthand for the measurable field of \(\Phi\)
  modulo the null-sets of \(\mu^{\mathcal{A,B,C}}_X(p)\).
\end{lemma}
\begin{lproof}
  By lemma~\ref{lemma:multiplication_instrument},
  \(\mu^{\mathcal{A,B,C}}_X(p)\) is a quantum instrument, hence there is a
  measure \(\nu_{\mu(p)}\) on \(X\) that dominates it. By assumption, \(p\)
  has the NEP, so there is a measure \(\nu_p\) on \(\monad X\) and a normal CPU
  map \(\overline{p} : L^\infty(\nu_p) \wtimes \mathcal{B} \to \mathcal{C}\)
  where \(p(E)(a) = \overline{p}(1_E \otimes a)\).

  For any bounded ultraweakly measurable map \(\Phi : \monad X \to
  \mathcal{A}\), consider the mapping \(\monad X \to \mathcal{B} : q \mapsto
  \overline{q}([\Phi]_q)\). This mapping is bounded by \(\norm{z}\) and
  measurable since \(\langle \phi, \overline{\operatorname{ev}}_\Phi(q)
  \rangle = \langle \phi, \overline{q}([\Phi]_q) \rangle\) is the
  composition of the measurable maps \(q \mapsto \overline{q}\) and
  \(\operatorname{ev}_{\phi,[\Phi]_q}\), so we can define
  \begin{equation}
    K(\Phi) = \overline{p}([q \mapsto \overline{\operatorname{ev}}_\Phi(q)]_p)
    = \overline{p}([q \mapsto \overline{q}([\Phi]_q)]_p) \in \mathcal{C}
  \end{equation}
  so that for simple functions \(s_{E,a}(x) = 1_{E}(x) \cdot a\),
  \begin{equation}
    K(s_{E,a}) = \overline{p}([q \mapsto q(E)(a)]_p) = \mu^{\mathcal{A,B,C}}_X(p)(E)(a).
  \end{equation}

  It remains to check that \(K\) depends only on the class
  \([\Phi]_{\mu(p)}\). Suppose first that \(\Phi\) is positive and vanishes
  outside a \(\nu_{\mu(p)}\)-null set \(N\). Then
  \begin{equation}
    0 \leqslant [\Phi]_q \leqslant \norm{\Phi} \cdot [1_N \otimes 1_A]_q
  \end{equation}
  for each \(q\), hence
  \begin{equation}
    0 \leqslant \overline{q}([\Phi]_q) \leqslant \norm{\Phi} \cdot q(N)(1_A).
  \end{equation}

  Applying \(\overline{p}\) gives
  \begin{equation}
    0 \leqslant K(\Phi)
    \leqslant \norm{\Phi} \cdot \overline{p}([q \mapsto q(N)(1_A)]_p)
    = \norm{\Phi} \cdot \mu_X^{A,B,C}(p)(N)(1_A) = 0,
  \end{equation}
  because \(\nu_{\mu(p)}\) dominates the instrument \(\mu_X^{A,B,C}(p)\). Hence \(K(\Phi)
  = 0\).

  By decomposing into positive and negative real and imaginary parts,
  the same conclusion holds for arbitrary bounded fields \(\Phi\)
  which vanish \(\nu_{\mu(p)}\)-almost everywhere. Thus \(K\) factors
  through the quotient by \(\nu_{\mu(p)}\). This defines the desired
  normal extension on the ultraweakly dense class of field elements as
  \(\overline{\mu^{\mathcal{A,B,C}}_X(p)}([\Phi]_{\mu(p)}) \coloneqq K(\Phi)\)
  and hence by normal continuation on all of \(L^\infty(\nu_{\mu(p)}) \otimes
  \mathcal{A}\).

  \proofsection{\(\overline{\mu^{\mathcal{A,B,C}}_X(p)}\) is unital}
  By lemma~\ref{lemma:multiplication_instrument},
  \begin{align}
    \overline{\mu^{\mathcal{A,B,C}}_X(p)}(1_X \otimes 1_\mathcal{A})
    &= \overline{\mu^{\mathcal{A,B,C}}_X(p)}([s_{X,1_\mathcal{A}}]_{\mu(p)}) \\
    &= \mu^{\mathcal{A,B,C}}_X(p)(X)(1_\mathcal{A}) = 1_\mathcal{C}.
  \end{align}

  \proofsection{\(\overline{\mu^{\mathcal{A,B,C}}_X(p)}\) is completely positive}
  Consider the matrix amplification \(\overline{\mu(p)}^{(n)}\)
  of \(\overline{\mu^{\mathcal{A,B,C}}_X(p)}\), which is
  uniquely defined by
  \begin{align}
    \overline{\mu(p)}^{(n)}([\Phi_{jk}]_{\mu(p)})
    &= [K(\Phi_{jk})] \\
    &= \big[\overline{p}([q \mapsto \overline{q}([\Phi_{jk}]_q)]_p)\big]_{jk} \\
    &= \overline{p}^{(n)}(\big[[q \mapsto \overline{q}([\Phi_{jk}]_q)]_p\big]_{jk})
  \end{align}
  and define an intermediate map \(T(\Phi) = [q \mapsto
  \overline{q}([\Phi]_q)]_p\), so that \(\overline{\mu^{\mathcal{A,B,C}}_X(p)}
  = \overline{p} \circ T\).

  Now, given the identification \(M_n(L^\infty(\nu_{\mu(p)}) \wtimes
  \mathcal{A}) \cong L^\infty(\nu_{\mu(p)}) \wtimes M_n(\mathcal{A})\), for
  any positive element of \(M_n(L^\infty(\nu_{\mu(p)}) \wtimes \mathcal{A})\)
  we can pick a bounded ultraweaky measurable representative \(\Phi : X \to
  M_n(\mathcal{A})_+\), which we write as \(\Phi(x) = [\Phi_{jk}(x)]_{jk}\).

    For each \(q \in \monad X\), this determines a field \([\Phi]_q \in
  M_n(L^\infty(\nu_{q}) \wtimes \mathcal{A}) \cong L^\infty(\nu_{q}) \wtimes
  M_n(\mathcal{A})\) which is also positive. Then we have that
  \begin{equation}
    T^{(n)}(\Phi) = [q \mapsto \overline{q}^{(n)}([\Phi]_q)]_p
  \end{equation}

  Assume that \(f \in L^1(\nu_p)\) and \(\phi \in \mathcal{B}_*\) are both
  positive, then
  \begin{align}
    \langle f \otimes \phi, T^{(n)}(\Phi) \rangle
    &= \langle f \otimes \phi, [q \mapsto \overline{q}^{(n)}([\Phi]_q)]_p \rangle \\
    &= \int_X f(x) \langle \phi, \overline{q}^{(n)}([\Phi]_q \rangle \ \nu_p(\dd{x})
  \end{align}
  which is immediately positive. Therefore, \(T^{(n)}\) is completely positive,
  and hence so is \(\overline{\mu^{\mathcal{A,B,C}}_X(p)}\).

  \proofsection{\(\overline{\mu^{\mathcal{A,B,C}}_X(p)}\) is normal}
  Let \([\Phi_\lambda]_{\mu(p)}\) be a bounded increasing net of
  positive elements in \(L^\infty(\nu_p) \wtimes \mathcal{A}\) with
  supremum \([\Phi]_{\mu(p)}\). For each \(q \in \monad X\), normality
  of \(\overline{q}\) gives \(\overline{q}([\Phi_\lambda]_q) \uparrow
  \overline{q}([\Phi]_q)\), then normality of \(\overline{p}\) gives
  \begin{equation}
    \overline{p}([q \mapsto \overline{q}([\Phi_\lambda]_q)]_p) \uparrow
    \overline{p}([q \mapsto \overline{q}([\Phi]_q)]_p)
  \end{equation}
  so that \(\overline{\mu(p)}\) is normal.

  \medskip

  We have therefore shown that the CPU map \(\overline{\mu(p)} : L^\infty(m)
  \wtimes \mathcal{A} \to \mathcal{C}\) is the normal extension of
  \(\mu_X^\mathcal{A,B,C}(p)\) which therefore has the NEP.
\end{lproof}

\begin{lemma}
  \(\mu^{\mathcal{A,B,C}}_X : \monad[B][C] \monad X  \to \monad[A][C] X \)
  is measurable.
\end{lemma}
\begin{lproof}
  Since each \(\monad\) is equipped with the evaluation \(\sigma\)-algebra,
  it suffices to show that \(\operatorname{ev}_{E,\phi,a} \circ
  \mu_X^{\mathcal{A,B,C}}\) is measurable for all \(E \in \sigma_X\), \(\phi
  \in \mathcal{C}_*\), and \(a \in \mathcal{A}\), i.e.\ that the map
  \begin{equation}
    \label{equation:multiplication_evaluation}
    p \longmapsto \operatorname{ev}_{E,\phi,a}\left(\textstyle\int_{\monad(X)} p(\dd{q}) \circ q\right)
  \end{equation}
  is measurable. Note that for any \(E \in \sigma_X\), \(a \in \mathcal{A}\),
  and \(\phi \in \mathcal{B}_*\),
  \begin{align}
    \operatorname{ev}_{\phi,1_E \otimes a}(\overline{\mu})
    = \langle \phi, \overline{\mu}(1_E \otimes a) \rangle
    = \langle \phi, \mu(E)(a) \rangle
    = \operatorname{ev}_{E,\phi,a}(\mu)
  \end{align}
  which by linearity extends to all simple functions \(s = \sum_{k=1}^n
  1_{E_k} \otimes a_k\) as \(\operatorname{ev}_{\phi,s}(\overline{\mu}) =
  \sum_{k=1}^n \operatorname{ev}_{E_k,\phi,a_k}(\mu)\). Since the pointwise
  limit of measurable functions is measurable, it follows that the map
  \(\monad X \to \CPSU[L^\infty(\nu) \wtimes A][B] : p \mapsto \overline{p}\)
  is measurable. Furthermore,
  \begin{align}
    \operatorname{ev}_{E,\phi,a}\left(\textstyle\int_{\monad(X)} p(\dd{q}) \circ q\right)
    &= \left\langle \phi, \textstyle\int_{\monad(X)} p(\dd{q}) \circ q(E)(a) \right\rangle \\
    &= \left\langle \phi, \overline{p}([q \mapsto q(E)(a)]_p) \right\rangle \\
    &= \operatorname{ev}_{\phi,[q \mapsto q(E)(a)]_p}(\overline{p}),
  \end{align}
  and so the mapping of equation~\eqref{equation:multiplication_evaluation}
  is measurable as the composition of measurable maps.
\end{lproof}

\begin{lemma}
  The family \((\mu^{\mathcal{A,B,C}}_X)_X\) describes a natural transformation
  \(\monad[B][C] \circ \monad \to \monad[A][C]\).
\end{lemma}
\begin{lproof}
  This is essentially immediate, and entirely analogous to the Giry case
  (up to tracking the indices \(\mathcal{A,B,C}\)).
\end{lproof}

\subsection{The monad laws}

\begin{proposition}
  For any \(\mathcal{A,B,C,D} \in \Wstar\) and \(X \in \meas\), the diagram
  \begin{equation}
    \begin{tikzcd}[column sep=2cm]
      \monad[C][D] \monad[B][C] \monad X & \monad[B][D] \monad X \\
      \monad[C][D] \monad[A][C] X & \monad[A][D] X
      \ar[from=1-1, to=1-2, "\mu^\mathcal{B,C,D}_{\monad X}"]
      \ar[from=1-1, to=2-1, "{\monad[C][D]} \mu^\mathcal{A,B,C}_X"']
      \ar[from=1-2, to=2-2, "\mu^\mathcal{A,B,D}_X"]
      \ar[from=2-1, to=2-2, "\mu^\mathcal{A,C,D}_X"]
    \end{tikzcd}
  \end{equation}
  commutes.
\end{proposition}
\begin{lproof}
  Let \(p \in \monad[C][D] \monad[B][C] \monad X\), then for any measurable
  \(E \subseteq X\) and \(a \in \mathcal{A}\),
  \begin{align}
    \mu^\mathcal{A,C,D}_X({\monad[C][D]} \mu^\mathcal{A,B,C}_X(p))(E)(a)
    &= \int_{\monad[A][C] X} {\monad[C][D]} \mu^\mathcal{A,B,C}_X(p)(\dd{q}) \circ q(E)(a) \\
    &= \int_{\monad[B][C] (\monad X)} p(\dd{q}) \circ \mu^\mathcal{A,B,C}_X(q)(E)(a) \\
    &= \overline{p}\left( \left[q \mapsto \mu^\mathcal{A,B,C}_X(q)(E)(a) \right]_p \right) \\
    &= \overline{p}\left( \left[q \mapsto \int_{\monad X} q(\dd{r}) \circ r(E)(a) \right]_p \right) \\
    &= \overline{p}\left( \left[q \mapsto \overline{q}([r \mapsto r(E)(a)]_q ) \right]_p \right) \\
    &= \overline{\mu^\mathcal{B,C,D}_{\monad X}(p)}([r \mapsto r(E)(a)]_{\mu(p)}) \label{eq:NEP_applied} \\
    &= \int_{\monad X} \mu^\mathcal{B,C,D}_{\monad X}(p)(\dd{r}) \circ r(E)(a) \\
    &= \mu^\mathcal{A,B,D}_X (\mu^\mathcal{B,C,D}_{\monad X}(p))(E)(a)
  \end{align}
  as required, where we have applied lemma~\ref{lemma:multiplication_NEP} to obtain equation~\eqref{eq:NEP_applied}.
\end{lproof}

\begin{proposition}
  For any \(\mathcal{A,B} \in \Wstar\) and \(X \in \meas\), the diagrams
  \begin{equation}
    \begin{tikzcd}
      \monad X & \monad \monad[A][A] X \\
      & \monad X
      \ar[from=1-1, to=1-2, "{\monad} \eta^\mathcal{A}_X "]
      \ar[from=1-2, to=2-2, "\mu_X^\mathcal{A,A,B}"]
      \ar[from=1-1, to=2-2, "1_{\monad X}"']
    \end{tikzcd}
    \qand
    \begin{tikzcd}
      \monad X & \monad[B][B] \monad X \\
      & \monad X
      \ar[from=1-1, to=1-2, "\eta^\mathcal{B}_{\monad X}"]
      \ar[from=1-2, to=2-2, "{\mu_X^\mathcal{A,B,B}}"]
      \ar[from=1-1, to=2-2, "1_{\monad X}"']
    \end{tikzcd}
  \end{equation}
  commute.
\end{proposition}
\begin{lproof}
  \proofsection{Left}
  Let \(p \in \monad X\), then for any measurable \(E \subseteq X\) and \(a
  \in \mathcal{A}\),
  \begin{align}
    \mu^\mathcal{A,A,B}_X(\monad \eta_X^\mathcal{A}(p))(E)(a)
    &= \int_{\monad[A][A] X} {\monad[A][B]} \eta_X^\mathcal{A}(p)(\dd{q}) \circ q(E)(a) \\
    &= \int_X p(\dd{x}) \circ \eta_X^\mathcal{A}(x)(E)(a) \\
    &= \overline{p}([\eta_X^\mathcal{A}(-)(E)(a)]_p) \\
    &= \overline{p}(1_E \otimes a) \\
    &= p(E)(a),
  \end{align}
  so \(\mu^\mathcal{A,A,B}_X(\monad \eta_X^\mathcal{A}(p)) = p\).

  \proofsection{Right}
  Let \(p \in \monad X\), then for any measurable \(E \subseteq
  X\) and \(a \in \mathcal{A}\),
  \begin{align}
    \mu^\mathcal{A,B,B}_X(\eta_{\monad X}^\mathcal{B}(p))(E)(a)
    &= \mu^\mathcal{A,B,B}_X(\delta_p^{\mathcal{B}})(E)(a) \\
    &= \int_{\monad X} \delta_p^{\mathcal{B}}(\dd{q}) \circ q(E)(a) \\
    &= \overline{\delta_p^{\mathcal{B}}}([q \mapsto q(E)(a)]_{\delta_p}) \\
    &= p(E)(a),
  \end{align}
  where we applied lemma~\ref{lemma:dirac_NEP} in the last step. It follows
  that \(\mu^\mathcal{A,B,B}_X(\eta_{\monad X}^\mathcal{B}(p)) = p\).
\end{lproof}

Putting all of these results together we have that:
\begin{theorem}
  \((Q,\mu,\eta)\) is a \(\Wstar\)-parameterised monad on \(\meas\).
\end{theorem}

In particular, we obtain immediate embeddings of classical probabilistic
processes:
\begin{corollary}
  \((Q_{\C,\C},\mu^{\C,\C,\C},\eta^\C)\) is the Giry monad.
\end{corollary}

\section{Quantum Markov kernels}
\label{section:kleisli}
Let \(\mathcal{A,B}\) be W*-algebras, and \((X,\sigma_X),(Y,\sigma_Y)\)
be measurable spaces. A \emph{quantum Markov kernel} \((X,\mathcal{A})
\to (Y,\mathcal{B})\) is a mapping \(\Phi : \sigma_X \times Y \to
\CPSU\) such that
\begin{itemize}
  \item for each \(y \in Y\), the mapping \(\sigma_X \to \CPSU : E \mapsto
  \Phi(E,y)\) is an NEP quantum instrument; and,
  \item for each measurable \(E \in \sigma_X\), the mapping \(Y \to \CPSU :
  y \mapsto \Phi(E,y)\) is a measurable map.
\end{itemize}

Following the quantum computing convention of drawing classical data as doubled
wires, we draw quantum Markov kernels \((X,\mathcal{A}) \to (Y,\mathcal{B})\)
as diagrams:
\begin{equation*}
  \begin{tikzpicture}
	\begin{pgfonlayer}{nodelayer}
		\node [style=none] (4) at (-1.375, 1) {};
		\node [style=none] (5) at (-1.375, -1) {};
		\node [style=none] (6) at (1.375, -1) {};
		\node [style=none] (7) at (1.375, 1) {};
		\node [style=none] (0) at (-0.625, 0.75) {};
		\node [style=none] (1) at (-0.625, -0.75) {};
		\node [style=none] (2) at (0.625, -0.75) {};
		\node [style=none] (3) at (0.625, 0.75) {};
		\node [style=none] (8) at (0.625, 0.475) {};
		\node [style=none] (12) at (1.375, 0.475) {};
		\node [style=none] (16) at (0, 0) {$\Phi$};
		\node [style=none] (42) at (-1.375, -0.375) {};
		\node [style=none] (46) at (-0.625, -0.375) {};
		\node [style=none] (60) at (0.625, 0.35) {};
		\node [style=none] (64) at (1.375, 0.35) {};
		\node [style=none] (87) at (0.625, -0.375) {};
		\node [style=none] (88) at (1.375, -0.375) {};
		\node [style=none] (89) at (-1.375, 0.475) {};
		\node [style=none] (90) at (-0.625, 0.475) {};
		\node [style=none] (91) at (-1.375, 0.35) {};
		\node [style=none] (92) at (-0.625, 0.35) {};
		\node [style=none] (93) at (-1.75, 0.375) {$X$};
		\node [style=none] (94) at (1.75, 0.375) {$Y$};
		\node [style=none] (95) at (-1.75, -0.375) {$\mathcal{A}$};
		\node [style=none] (96) at (1.75, -0.375) {$\mathcal{B}$};
	\end{pgfonlayer}
	\begin{pgfonlayer}{edgelayer}
		\draw [style=background] (5.center)
			 to (4.center)
			 to (7.center)
			 to (6.center)
			 to cycle;
		\draw [style=bbox] (1.center)
			 to (0.center)
			 to (3.center)
			 to (2.center)
			 to cycle;
		\draw (8.center) to (12.center);
		\draw (42.center) to (46.center);
		\draw (60.center) to (64.center);
		\draw (87.center) to (88.center);
		\draw (89.center) to (90.center);
		\draw (91.center) to (92.center);
	\end{pgfonlayer}
\end{tikzpicture}
,
  \qq{so a quantum instrument is drawn}
  \begin{tikzpicture}
	\begin{pgfonlayer}{nodelayer}
		\node [style=none] (4) at (-1.375, 1) {};
		\node [style=none] (5) at (-1.375, -1) {};
		\node [style=none] (6) at (1.375, -1) {};
		\node [style=none] (7) at (1.375, 1) {};
		\node [style=none] (0) at (-0.625, 0.75) {};
		\node [style=none] (1) at (-0.625, -0.75) {};
		\node [style=none] (2) at (0.625, -0.75) {};
		\node [style=none] (3) at (0.625, 0.75) {};
		\node [style=none] (8) at (-1.375, 0.475) {};
		\node [style=none] (12) at (-0.625, 0.475) {};
		\node [style=none] (16) at (0, 0) {$q$};
		\node [style=none] (42) at (-1.375, -0.375) {};
		\node [style=none] (46) at (-0.625, -0.375) {};
		\node [style=none] (60) at (-1.375, 0.35) {};
		\node [style=none] (64) at (-0.625, 0.35) {};
		\node [style=none] (87) at (0.625, -0.375) {};
		\node [style=none] (88) at (1.375, -0.375) {};
		\node [style=none] (89) at (-1.875, 0.375) {$X$};
		\node [style=none] (91) at (-1.75, -0.375) {$\mathcal{A}$};
		\node [style=none] (92) at (1.75, -0.375) {$\mathcal{B}$};
	\end{pgfonlayer}
	\begin{pgfonlayer}{edgelayer}
		\draw [style=background] (5.center)
			 to (4.center)
			 to (7.center)
			 to (6.center)
			 to cycle;
		\draw [style=bbox] (1.center)
			 to (0.center)
			 to (3.center)
			 to (2.center)
			 to cycle;
		\draw (8.center) to (12.center);
		\draw (42.center) to (46.center);
		\draw (60.center) to (64.center);
		\draw (87.center) to (88.center);
	\end{pgfonlayer}
\end{tikzpicture}
.
\end{equation*}

\begin{example}
  If \(X\) and \(Y\) are discrete finite measurable spaces, then a quantum
  Markov kernel \((X,\mathcal{A}) \to (Y,\mathcal{B})\) can equivalently be
  seen as a \(\abs{Y} \times \abs{X}\) matrix who elements are taken from
  \(\CPSU\).
\end{example}

We can straightforwardly curry kernels to instrument-valued functions and
back:
\begin{proposition}
  There is a bijection between quantum Markov kernels \((X,\mathcal{A}) \to
  (Y,\mathcal{B})\) and Kleisli maps \(X \to \monad Y\).
\end{proposition}
We can therefore obtain the composition of quantum Markov kernels as the
composition of the corresponding Kleisli maps:
\begin{equation*}
  \begin{tikzpicture}
	\begin{pgfonlayer}{nodelayer}
		\node [style=none] (4) at (-7.125, 1.25) {};
		\node [style=none] (5) at (-7.125, -1.25) {};
		\node [style=none] (6) at (-1.375, -1.25) {};
		\node [style=none] (7) at (-1.375, 1.25) {};
		\node [style=none] (0) at (-6.375, 0.75) {};
		\node [style=none] (1) at (-6.375, -0.75) {};
		\node [style=none] (2) at (-5.125, -0.75) {};
		\node [style=none] (3) at (-5.125, 0.75) {};
		\node [style=none] (8) at (-5.125, 0.475) {};
		\node [style=none] (12) at (-3.375, 0.475) {};
		\node [style=none] (16) at (-5.75, 0) {$\Psi$};
		\node [style=none] (42) at (-7.125, -0.375) {};
		\node [style=none] (46) at (-6.375, -0.375) {};
		\node [style=none] (60) at (-5.125, 0.35) {};
		\node [style=none] (64) at (-3.375, 0.35) {};
		\node [style=none] (87) at (-5.125, -0.375) {};
		\node [style=none] (88) at (-3.375, -0.375) {};
		\node [style=none] (89) at (-7.125, 0.475) {};
		\node [style=none] (90) at (-6.375, 0.475) {};
		\node [style=none] (91) at (-7.125, 0.35) {};
		\node [style=none] (92) at (-6.375, 0.35) {};
		\node [style=none] (93) at (-7.5, 0.375) {$X$};
		\node [style=none] (94) at (-4.25, 0.875) {$Y$};
		\node [style=none] (95) at (-7.5, -0.375) {$\mathcal{A}$};
		\node [style=none] (96) at (-4.25, -0.7) {$\mathcal{B}$};
		\node [style=none] (97) at (1.375, 1) {};
		\node [style=none] (98) at (1.375, -1) {};
		\node [style=none] (99) at (5.625, -1) {};
		\node [style=none] (100) at (5.625, 1) {};
		\node [style=none] (101) at (2.125, 0.75) {};
		\node [style=none] (102) at (2.125, -0.75) {};
		\node [style=none] (103) at (4.875, -0.75) {};
		\node [style=none] (104) at (4.875, 0.75) {};
		\node [style=none] (105) at (4.875, 0.475) {};
		\node [style=none] (106) at (5.625, 0.475) {};
		\node [style=none] (107) at (3.5, 0) {$\Phi \odot \Psi$};
		\node [style=none] (108) at (1.375, -0.375) {};
		\node [style=none] (109) at (2.125, -0.375) {};
		\node [style=none] (110) at (4.875, 0.35) {};
		\node [style=none] (111) at (5.625, 0.35) {};
		\node [style=none] (112) at (4.875, -0.375) {};
		\node [style=none] (113) at (5.625, -0.375) {};
		\node [style=none] (114) at (1.375, 0.475) {};
		\node [style=none] (115) at (2.125, 0.475) {};
		\node [style=none] (116) at (1.375, 0.35) {};
		\node [style=none] (117) at (2.125, 0.35) {};
		\node [style=none] (118) at (1, 0.375) {$X$};
		\node [style=none] (119) at (6, 0.375) {$Z$};
		\node [style=none] (120) at (1, -0.375) {$\mathcal{A}$};
		\node [style=none] (121) at (6, -0.375) {$\mathcal{C}$};
		\node [style=none] (122) at (-3.375, 0.75) {};
		\node [style=none] (123) at (-3.375, -0.75) {};
		\node [style=none] (124) at (-2.125, -0.75) {};
		\node [style=none] (125) at (-2.125, 0.75) {};
		\node [style=none] (126) at (-2.125, 0.475) {};
		\node [style=none] (127) at (-1.375, 0.475) {};
		\node [style=none] (128) at (-2.75, 0) {$\Phi$};
		\node [style=none] (131) at (-2.125, 0.35) {};
		\node [style=none] (132) at (-1.375, 0.35) {};
		\node [style=none] (133) at (-2.125, -0.375) {};
		\node [style=none] (134) at (-1.375, -0.375) {};
		\node [style=none] (140) at (-1, 0.375) {$Z$};
		\node [style=none] (142) at (-1, -0.375) {$\mathcal{C}$};
		\node [style=none] (143) at (0, 0) {$=$};
	\end{pgfonlayer}
	\begin{pgfonlayer}{edgelayer}
		\draw [style=background] (5.center)
			 to (4.center)
			 to (7.center)
			 to (6.center)
			 to cycle;
		\draw [style=bbox] (1.center)
			 to (0.center)
			 to (3.center)
			 to (2.center)
			 to cycle;
		\draw (8.center) to (12.center);
		\draw (42.center) to (46.center);
		\draw (60.center) to (64.center);
		\draw (87.center) to (88.center);
		\draw (89.center) to (90.center);
		\draw (91.center) to (92.center);
		\draw [style=background] (98.center)
			 to (97.center)
			 to (100.center)
			 to (99.center)
			 to cycle;
		\draw [style=bbox] (102.center)
			 to (101.center)
			 to (104.center)
			 to (103.center)
			 to cycle;
		\draw (105.center) to (106.center);
		\draw (108.center) to (109.center);
		\draw (110.center) to (111.center);
		\draw (112.center) to (113.center);
		\draw (114.center) to (115.center);
		\draw (116.center) to (117.center);
		\draw [style=bbox] (123.center)
			 to (122.center)
			 to (125.center)
			 to (124.center)
			 to cycle;
		\draw (126.center) to (127.center);
		\draw (131.center) to (132.center);
		\draw (133.center) to (134.center);
	\end{pgfonlayer}
\end{tikzpicture}

\end{equation*}
where explicitly,
\begin{equation}
  \Phi \odot \Psi(E,x)(a) = \int_Y \Psi(\dd{y},z) \circ \Phi(E,y)(a),
\end{equation}
for any \(x \in X\), \(E \in \sigma_Z\), and \(a \in \mathcal{A}\).

{\raggedright\printbibliography}

\appendix

\section{Monads and parameterised monads}
\label{appendix:monads}
We recall the definitions of monads, Kleisli categories, and parameterised
monads. We only overview parameterised monads here, and refer the interested
reader directly to \cite{atkey:parameterised_monads} of details.

\subsection{Monads}

\begin{definition}
  A \emph{monad} on a category \(\mathsf C\) consists of an endofunctor \(T :
  \mathsf C \to \mathsf C\) together with natural transformations \(\eta :
  \mathrm{id}_{\mathsf C} \Rightarrow T\) and \(\mu : T^2 \Rightarrow T\),
  called the \emph{unit} and \emph{multiplication}, such that the following
  diagrams commute:
  \begin{equation}
    \begin{tikzcd}
    T^3 X \ar[r, "T\mu_X"] \ar[d, "\mu_{T X}"']
    & T^2 X \ar[d, "\mu_X"] \\
    T^2 X \ar[r, "\mu_X"']
    & T X
    \end{tikzcd}
  \end{equation}
  and
  \begin{equation}
    \begin{tikzcd}
    T X \ar[r, "\eta_{T X}"] \ar[dr, "\mathrm{id}_{T X}"']
    & T^2 X \ar[d, "\mu_X"] &
    T X \ar[l, "T\eta_X"'] \ar[dl, "\mathrm{id}_{T X}"] \\
    & T X &
    \end{tikzcd}
  \end{equation}
  for every object \(X \in \mathsf C\).
\end{definition}

\subsection{Kleisli composition}

A monad packages a notion of composition for morphisms of the form \(X \to
T Y\), called \emph{Kleisli morphisms}.

Given morphisms
\begin{equation}
  f : X \to T Y,
  \qand
  g : Y \to T Z,
\end{equation}
their \emph{Kleisli composite} is the morphism
\begin{equation}
  g \odot f
  :=
  X \xrightarrow{f} T Y
  \xrightarrow{T g} T^2 Z
  \xrightarrow{\mu_Z} T Z.
\end{equation}
The unit at \(X\), \(\eta_X : X \to T X\), acts as the identity Kleisli
morphism on \(X\): for every \(f : X \to T Y\),
\begin{equation}
  f \odot \eta_X = f,
  \qand
  \eta_Y \odot f = f.
\end{equation}
With this structure, the Kleisli morphisms assemble into
a category with the same objects as \(\mathsf{C}\), called the \emph{Kleisli
category} \(\mathsf{C}_T\) of the monad \(T\).

\subsection{Parameterised monads}

Parameterised monads are an indexed version of monads in which processes carry
both an input parameter and an output parameter. We state the definition for
a parameter category \(\mathsf P\). In the special case where \(\mathsf P\)
is discrete, this reduces to a family indexed by a set.

A \emph{parameterised monad} on \(\mathsf C\) indexed by a category
\(\mathsf P\) consists of a functor
\[
  T : \mathsf P^{\mathrm{op}} \times \mathsf P \to [\mathsf C,\mathsf C],
\]
written \((p,q) \mapsto T^{p,q}\), together with natural transformations
\(
  \eta_p : \mathrm{id}_{\mathsf C} \Rightarrow T^{p,p}
\)
for each \(p \in \mathsf P\), and
\(
  \mu^{p,q,r} : T^{p,q} \circ T^{q,r} \Rightarrow T^{p,r}
\)
for each composable triple \(p,q,r \in \mathsf P\), satisfying the
associativity and unit axioms below.

For every object \(X \in \mathsf C\), the multiplication has components
\(
  \mu^{p,q,r}_X : T^{p,q}(T^{q,r} X) \to T^{p,r} X.
\)
For every \(X \in \mathsf C\), the diagram
\[
  \begin{tikzcd}[column sep=2cm]
    T^{p,q} T^{q,r} T^{r,s} X
    \ar[r, "T^{p,q}(\mu^{q,r,s}_{X})"]
    \ar[d, "\mu^{p,q,r}_{T^{r,s}X}"']
    & T^{p,q} T^{q,s} X
    \ar[d, "\mu^{p,q,s}_{X}"] \\
    T^{p,r} T^{r,s} X
    \ar[r, "\mu^{p,r,s}_{X}"']
    & T^{p,s} X
  \end{tikzcd}
\]
commutes.

The unit axioms say that, for all \(p,q \in \mathsf P\), the diagrams
\begin{equation}
  \begin{tikzcd}
    T^{p,q} X & T^{p,q} T^{q,q} X \\
    & \monad X
    \ar[from=1-1, to=1-2, "{T^{p,q}} \eta^q_X "]
    \ar[from=1-2, to=2-2, "\mu_X^{q,q,p}"]
    \ar[from=1-1, to=2-2, "1_{T{q,p} X}"']
  \end{tikzcd}
  \qand
  \begin{tikzcd}
    T^{p,q} X & T^{p,p} T^{q,p} X \\
    & T^{p,q} X
    \ar[from=1-1, to=1-2, "\eta^q_{T^{p,q} X}"]
    \ar[from=1-2, to=2-2, "{\mu_X^{q,p,p}}"]
    \ar[from=1-1, to=2-2, "1_{T^{p,q} X}"']
  \end{tikzcd}
\end{equation}
commute.

\subsection{Parameterised Kleisli composition}

Given a parameterised monad \(T\), a parameterised Kleisli morphism from \(X\)
to \(Y\), changing parameter \(p\) to parameter \(q\), is a morphism
\[
  f : X \to T^{p,q} Y
\]
in \(\mathsf C\).

If
\[
  f : X \to T^{p,q} Y,
  \qquad
  g : Y \to T^{q,r} Z,
\]
then their parameterised Kleisli composite is
\[
  g \odot f
  :=
  X \xrightarrow{f} T^{p,q} Y
  \xrightarrow{T^{p,q} g} T^{p,q} T^{q,r} Z
  \xrightarrow{\mu^{p,q,r}_Z} T^{p,r} Z.
\]
This composite is defined only when the output parameter \(q\) of \(f\)
agrees with the input parameter \(q\) of \(g\).
\begin{definition}
  The \emph{parameterised Kleisli category} associated to a parameterised monad
  \(T\) is the category \(\mathsf C_T\) defined as follows:
  \begin{itemize}
    \item objects are pairs \((p,X)\), where \(p \in \mathsf P\) and \(X
    \in \mathsf C\);
    \item a morphism
    \(
      (p,X) \to (q,Y)
    \)
    is a morphism
    \(
      X \to T^{p,q}Y
    \)
    in \(\mathsf C\);
    \item the identity on \((p,X)\) is
    \(
      \eta^p_{X} : X \to T^{p,p}X;
    \)
    \item composition is parameterised Kleisli composition.
  \end{itemize}
\end{definition}

The associativity axiom for \(\mu\) implies associativity of parameterised
Kleisli composition. The unit axioms imply that \(\eta_p\) acts as the identity
at parameter \(p\). Hence \(\mathsf C_T\) is indeed a category.
 
\end{document}